\documentclass[reprint,showpacs,twocolumn,aps,superscriptaddress]{revtex4-2}
\usepackage{amsfonts}
\usepackage{amsmath}
\usepackage{amssymb}
\usepackage{graphicx}
\usepackage{verbatim}
\usepackage{textcomp}
\usepackage{hyperref}
\usepackage{upgreek}

\providecommand{\U}[1]{\protect\rule{.1in}{.1in}}

\begin{document}
\title{High sensitivity air-coupled MHz frequency ultrasound detection using on-chip microcavities}

\author{Hao Yang}
\affiliation{Beijing National Laboratory for Condensed Matter Physics, Institute of Physics, Chinese Academy of Sciences, Beijing 100094, P. R. China}
\affiliation{University of Chinese Academy of Sciences, Beijing 100049, P. R. China}
\author{Zhi-Gang Hu}
\affiliation{Beijing National Laboratory for Condensed Matter Physics, Institute of Physics, Chinese Academy of Sciences, Beijing 100094, P. R. China}
\affiliation{University of Chinese Academy of Sciences, Beijing 100049, P. R. China}
\author{Yuechen Lei}
\affiliation{Beijing National Laboratory for Condensed Matter Physics, Institute of Physics, Chinese Academy of Sciences, Beijing 100094, P. R. China}
\affiliation{University of Chinese Academy of Sciences, Beijing 100049, P. R. China}
\author{Xuening Cao}
\affiliation{Beijing National Laboratory for Condensed Matter Physics, Institute of Physics, Chinese Academy of Sciences, Beijing 100094, P. R. China}
\affiliation{University of Chinese Academy of Sciences, Beijing 100049, P. R. China}
\author{Min Wang}
\affiliation{Beijing National Laboratory for Condensed Matter Physics, Institute of Physics, Chinese Academy of Sciences, Beijing 100094, P. R. China}
\affiliation{University of Chinese Academy of Sciences, Beijing 100049, P. R. China}
\author{Jialve Sun}
\affiliation{College of Future Technology, Peking University, Beijing 100871, China}
\author{Zhanchun Zuo}
\affiliation{Beijing National Laboratory for Condensed Matter Physics, Institute of Physics, Chinese Academy of Sciences, Beijing 100094, P. R. China}
\affiliation{Songshan Lake Materials Laboratory, Dongguan 523808, Guangdong, P. R. China}
\author{Changhui Li}
\affiliation{College of Future Technology, Peking University, Beijing 100871, China}
\author{Xiulai Xu}
\affiliation{Beijing National Laboratory for Condensed Matter Physics, Institute of Physics, Chinese Academy of Sciences, Beijing 100094, P. R. China}
\author{Bei-Bei Li}
\email{libeibei@iphy.ac.cn}
\affiliation{Beijing National Laboratory for Condensed Matter Physics, Institute of Physics, Chinese Academy of Sciences, Beijing 100094, P. R. China}
\affiliation{Songshan Lake Materials Laboratory, Dongguan 523808, Guangdong, P. R. China}

\date{\today}

\begin{abstract}

Owing to their dual-resonance enhanced sensitivity, cavity optomechanical systems provide an ideal platform for ultrasound sensing. In this work, we realize high sensitivity air-coupled ultrasound sensing from kilohertz (kHz) to megahertz (MHz) frequency range based on whispering gallery mode microcavities. Using a 57~\textmu m-diameter microtoroid with high optical \emph{Q} factor ($\sim$10$^7$) and mechanical \emph{Q} factor ($\sim$700), we achieve sensitivities of 46~\textmu Pa Hz$^{-1/2}$-10~mPa Hz$^{-1/2}$ in a frequency range of 0.25-3.2 MHz. Thermal-noise-limited sensitivity is realized around the mechanical resonance at 2.56~MHz, in a frequency range of 0.6~MHz. We also observe the second- and third-order mechanical sidebands, and quantitatively study the intensities of each mechanical sideband as a function of the mechanical displacement. Measuring the combination of signal to noise ratios at all sidebands has the potential to extend the dynamic range of ultrasound sensing. In addition, to improve the ultrasound sensitivity in the kHz frequency range, we use a microdisk with a diameter of 200~\textmu m, and achieve sensitivities of 1.83~\textmu Pa Hz$^{-1/2}$-10.4~mPa Hz$^{-1/2}$ in 30~kHz-1.65~MHz range. 

\end{abstract}

\maketitle

\section{Introduction}
Miniaturized high-sensitivity ultrasound sensors are key components in various applications, such as medical diagnostics \cite{1}, photoacoustic imaging and spectroscopy \cite{2,3,4}, nondestructive testing \cite{5}, sonar \cite{6,7}, trace gas monitoring \cite{8}, etc. Currently, the piezoelectric transducers are mostly widely used, but their sensitivities drop quickly when the size becomes smaller, leading to a typical sensor size of millimeter to centimeter scale \cite{9,10,11}. In order to realize both high sensitivity and spatial resolution, photonic ultrasound sensors that can be microfabricated on a silicon chip has been developed. Among them, cavity optomechanical systems \cite{12,13,14,15} attract increasing interest owing to their superior characteristics of high sensitivity, broad bandwidth, low power consumption, and chip-scale integration. In cavity optomechanical systems, displacement of the cavity can be optically read out via the optomechanical coupling. As the response is enhanced by the mechanical resonance, and the readout sensitivity can be also enhanced by the optical resonance, cavity optomechanical systems have been proved to be an ideal sensing platform for displacement \cite{16,17}, mass \cite{18,19,20}, force \cite{21,22,23}, acceleration \cite{24,25}, magnetic field \cite{26,27,28,29,30,31,32}, acoustic wave \cite{33,34,35,36,37,38,39,40,add toroid,41,42,43,44,45,46,47}, etc.

Ultrasound sensing using cavity optomechanical systems in the liquid environment have been demonstrated with various microcavity systems. Polymer materials are generally soft and can be easily deformed by acoustic waves, and therefore provide large sensing signals. Various polymer microcavities such as polystyrene \cite{33,34}, SU8 \cite{35} and polydimethylsiloxane (PDMS) \cite{36}, have been fabricated for ultrasound sensing, and achieved sensitivity at Pascal level and high bandwidth of tens-to-hundreds of MHz. A Fabry-Perot cavity has been fabricated at the end of an optical fiber using UV curable epoxy, which has realized a sensitivity of mPa Hz$^{-1/2}$ at the tens of MHz frequency range \cite{37}. Silicon microcavities have also attracted increasing interest, as they can be massively produced on a chip and their fabrication techniques have been well developed in the past few decades. Recently, Shnaiderman et al. demonstrated miniaturized high-sensitivity ultrasound sensing using an array of point like silicon waveguide-etalon detector on a silicon on insulator (SOI) platform, and realized a sensing bandwidth of hundreds of MHz \cite{38}. Later, Westerveld et al. demonstrated an optomechanical ultrasound sensor using a silicon microring cavity coupled with a thin film, with a 15~nm gap in between, and realized a sensitivity of mPa Hz$^{-1/2}$ in the tens of MHz frequency range \cite{39}. Silica microcavities have also been extensively explored for ultrasound sensing, due to their ultrahigh optical \emph{Q} factors. Ultrasound sensing in a liquid environment has been demonstrated using microtoroid \cite{40,add toroid} and microsphere \cite{34,41} cavities.

Air-coupled ultrasound sensing has specific applications such as gas photoacoustic spectroscopy \cite{48}, and non-contact ultrasonic medical imaging \cite{49}. Due to the large impedance mismatch at the acoustic source/air interface and the absorption loss of ultrasonic waves, air-coupled ultrasound detection requires ultrahigh sensitivity. Ultrasound sensing in air has been demonstrated using microbottle cavities, with sensitivities on the order of mPa Hz$^{-1/2}$ at tens to hundreds of kHz frequency range \cite{42,43}. Through detecting the acoustic wave induced modulation of the Brillouin laser in a microsphere, acoustic sensitivity of 267~\textmu Pa Hz$^{-1/2}$ has been realized in the kHz frequency range \cite{44}. Basiri-Esfahani et al. have realized ultrasound sensing in the thermal-noise-dominant regime using a spoked microdisk cavity, and achieved sensitivities of 8-300~\textmu Pa Hz$^{-1/2}$ in the frequency range from 1~kHz to 1~MHz \cite{45}. Up to now, high-sensitivity ultrasound sensing in air above 1~MHz has not been reported yet. 

In this work, we demonstrate an air-coupled ultrasound detection in the kHz-MHz frequency range, using on-chip microcavities. To extend the sensing frequency into the MHz range, we use a microtoroid with a diameter of 57~\textmu m which supports mechanical resonances in the MHz range. To decrease the constraint of the mechanical motion from the substrate, we use a two-step etching process to make a microtoroid with a thin silicon pedestal. This allows a high mechanical \emph{Q} factor of $\sim$700 of the first-order flapping mode at 2.56~MHz. Compared with the spoked microdisk used in Ref. \cite{45}, the microtoroid has less squeeze-film damping due to the larger allowed undercut (distance between the toroid and substrate). Combining with the high optical \emph{Q} factor of $\sim$10$^7$, thermal-noise-limited sensitivity has been reached in air, within a frequency range of 0.6~MHz. We have achieved sensitivities of 46~\textmu Pa Hz$^{-1/2}$-10~mPa Hz$^{-1/2}$ in the frequency range of 0.25-3.2~MHz. We have also observed second- and third-order mechanical sidebands in the noise power spectrum, when driving the sensor with an ultrasonic wave at the mechanical resonance, which is caused by the transduction nonlinearity. We measure the signal-to-noise ratios (SNRs) under different ultrasound pressures (\emph{P}), and find that $\sqrt{\rm{SNR}}$s of the first-, second-, and third-order mechanical sidebands is approximately proportional to \emph{P}, \emph{P}$^2$, and \emph{P}$^3$, respectively, which agree well with our theoretical results. In addition, in order to further improve the ultrasound sensitivity in the kHz frequency range, we use a microdisk with a diameter of 200~\textmu m which supports multiple mechanical resonances at hundreds of kHz range. The achieved sensitivities are 1.83~\textmu Pa Hz$^{-1/2}$-10.4~mPa Hz$^{-1/2}$ in 30~kHz-1.65~MHz range. The peak sensitivity of 1.83~\textmu Pa Hz$^{-1/2}$ at 189~kHz is the record sensitivity of cavity optomechanical ultrasound sensors reported so far. Our measured result shows that the sensitivities in the 30-700~kHz range are enhanced by the mechanical resonances of the tapered fiber we use to couple light into the microcavities.

\section{Methods}

\begin{figure}[t!]
	\begin{center}
		\includegraphics[width=8.6cm]{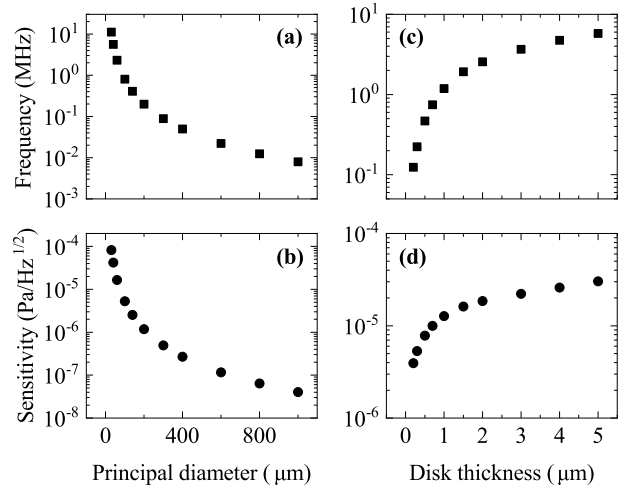}
	\end{center}
	\caption{(a) and (b) Simulated mechanical resonance frequency and calculated sensitivity of the sensor, with different principal diameters. (c) and (d) Simulated mechanical resonance frequency and calculated sensitivity of the sensor, with different disk thicknesses.}
	\label{fig1}
\end{figure}

The ultrasound sensitivity is determined by the noise of the sensor. In our detection system, the main sources of noise are laser noise from the probe light and mechanical thermal noise from the non-zero temperature environment. The laser noise mainly consists of the classical noise (including intensity noise and phase noise) which is the dominant noise source in the low frequency range, and the quantum shot noise which is dominant at high frequencies. For a microcavity with optical energy decay rate of $\kappa$ and mechanical damping rate of $\gamma$, the corresponding noise equivalent pressure, i.e., the ultrasound sensitivity $P_{\rm{min}}$, can be expressed as Eq. (1) \cite{13,45}
\begin{align}
P_{\rm{min}}(\omega)&  =\frac{1}{r\zeta A}\sqrt{\frac{S_{xx}^{\rm{shot}}+S_{xx}^{\rm{classical}}}{|\chi_{\rm{m}}|^2}+S_{FF}^{\rm{thermal}}}     \nonumber     \\
&   = \frac{1}{r\zeta A}\sqrt{\frac{\frac{\kappa(1+4\omega^2/\kappa^2)}{16 \eta n_{\rm{c}} G^2}+S_{xx}^{\rm{classical}}}{|\chi_{\rm{m}}|^2}+2m\gamma k_{\rm{B}} T}
\end{align}

where \emph{r} is the ratio of the pressure difference between the upper and lower surfaces of the device to the peak pressure at the antinode of the ultrasonic wave, as the cavity only moves by feeling the pressure difference between the upper and lower surfaces. $\zeta$ is the spatial overlap between the incident ultrasound and the mechanical displacement profile of the sensor. $\omega$ is the angular frequency of the incident ultrasonic wave, and \emph{A} is the sensor area. The first term under the square root denotes the optical noise, with $S_{xx}^{\rm{shot}}$ and $S_{xx}^{\rm{classical}}$ representing displacement power spectral densities of shot noise \cite{shot} and classical noise, respectively. $\eta$ is the total detection efficiency of light, and $n_{\rm{c}}$ is the number of photons in the cavity. \emph{G} = d$\omega$/d\emph{x} denotes the optomechanical coupling coefficient, quantifying the cavity frequency shift for unit mechanical displacement \emph{x}. The second term under the square-root quantifies thermal noise at temperature \emph{T}, introduced by both the intrinsic damping of the mechanical resonator and collisions with the gas molecules around the sensor. Here \emph{m} is the effective mass of the sensor, and $\chi_{\rm{m}}$($\omega$) is the mechanical susceptibility, quantified by $\chi_{\rm{m}}$($\omega$) = $\frac{1}{\emph{m}(\Omega^{2}-\omega^{2}-\emph{i}\gamma\omega)}$, with $\Omega$ being the angular frequency of the mechanical resonance. From the Eq. (1), we can see that the sensitivity is fundamentally limited by the thermal noise, if the measurement strength is strong enough to enable thermal noise dominating laser noise. As a result, reaching thermal-noise-limited regime is beneficial to achieving a better sensitivity. This can be realized by increasing the optical \emph{Q} factor, mechanical \emph{Q} factor, and the optomechanical coupling coefficient $G$. A larger sensing bandwidth can be obtained by increasing the thermal-noise-dominant frequency range.

The first-order flapping mode has a large spatial overlap with the ultrasonic wave coming from the top of the sensor, which is beneficial to achieving a good ultrasound sensitivity. We then optimize the ultrasound sensitivity for this mode, by changing the geometric parameters of the toroid. We first simulate the resonance frequency for different principal diameters of the cavity from 30 to 1000~\textmu m, with the result shown in Fig. \ref{fig1}(a). In the simulation, we keep the minor diameter of the toroid to be 6~\textmu m, and the disk thickness to be 2~\textmu m. It can be seen that, with the increase of the principal diameter, the resonance frequency decreases monotonously. We then calculate the thermal-noise-limited ultrasound sensitivity for different principal diameters of the toroid, as shown in Fig. \ref{fig1}(b). In the calculation, we use the mechanical \emph{Q} factor of $Q_m$ = 700, obtained from our experiment. It can be seen that with the increase of the principal diameter, the sensitivity gets better, due to the increased sensing area. Thus, for ultrasound sensing at high frequency range, a microcavity with a smaller diameter (and therefore higher mechanical frequency) is desired. While for low frequency ultrasound sensing, using a cavity with a larger diameter is beneficial to achieving a better sensitivity. We then simulate the resonance frequency and calculate the corresponding thermal-noise-limited sensitivity for different disk thicknesses from 200~nm to 5~\textmu m, as shown in Figs. \ref{fig1}(c) and \ref{fig1}(d), respectively. It can be seen that, with the disk thickness increases, the resonance frequency increases, and the sensitivity gets worse. 

In our experiment, in order to facilitate the fabrication of high optical \emph{Q} toroid, we choose the disk thickness to be 2~\textmu m instead of a thinner one. In order to optimize the ultrasound sensitivity at MHz frequency range, we use a toroid with a principal diameter of $\sim$57~\textmu m, minor diameter of $\sim$6~\textmu m, whose mechanical resonance frequency of the first-order flapping mode is $\sim$2.56~MHz and corresponding effective mass is $m=$ 12.1~ng. For an ideal case, $\zeta$ = 1, \emph{r} = 1, the corresponding sensitivity is calculated to be $\sim$18.5~\textmu Pa Hz$^{-1/2}$. In addition, to improve the sensitivity in the kHz frequency range, we use a microdisk with a diameter of 200~\textmu m to perform the ultrasound sensing. The mechanical resonance of the first-order flapping mode is 210~kHz, with an effective mass of 60.5~ng. The sensitivity of the mode is calculated to be $\sim$2.2~\textmu Pa Hz$^{-1/2}$ for $\zeta$ = 1, \emph{r} = 1, and $Q_m$ = 130 (obtained from the experiment).

\begin{figure}[t!]
\begin{center}
\includegraphics[width=8.6cm]{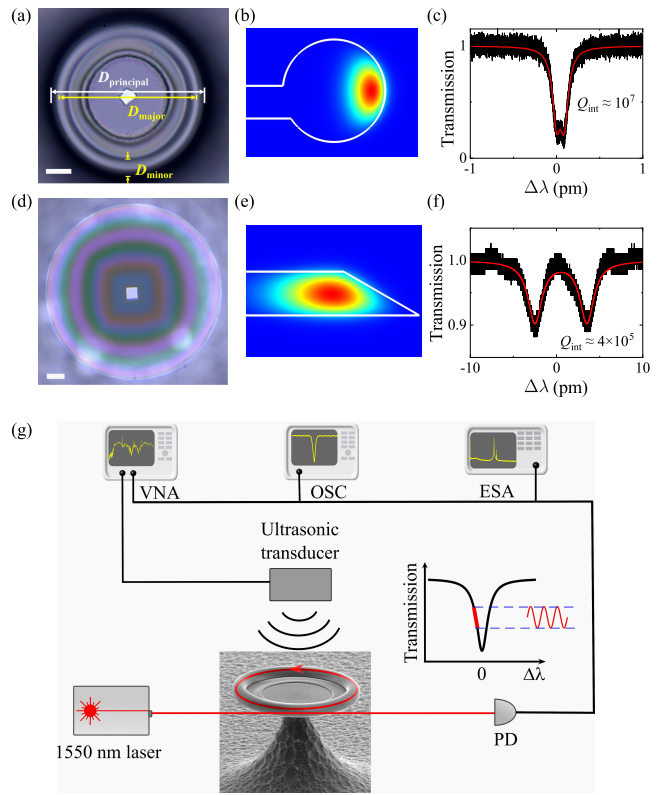}
\end{center}
\caption{(a) and (d) Top-view optical microscope images of the microtoroid and microdisk. In (a), \emph{D}$_{\rm{principal}}$, \emph{D}$_{\rm{major}}$ and \emph{D}$_{\rm{minor}}$ denote the principal diameter, major diameter, and minor diameter of the microtoroid, with \emph{D}$_{\rm{principal}}$ = \emph{D}$_{\rm{major}}$ + \emph{D}$_{\rm{minor}}$. The scale bar corresponds to 10~\textmu m in (a) and 20~\textmu m in (d). (b) and (e) Simulated optical field distributions of the fundamental WGM of the microtoroid and microdisk. (c) and (f) Transmission spectra of the microtoroid and microdisk around 1550~nm, with the red solid curves showing the double Lorentzian fitting results, from which we can obtain \emph{Q}$_{\rm{int}} \approx 10^7$ for the microtoroid and \emph{Q}$_{\rm{int}} \approx 4\times10^5$ for the microdisk. (g) Schematic diagram of the experimental setup for ultrasound sensing, with the inset showing the principle of ultrasound detection. PD, photodetector; VNA, vector network analyzer; OSC, oscilloscope; ESA, electronic spectrum analyzer. An SEM image of the microtoroid is shown in the setup.}
\label{fig2}
\end{figure}

The microtoroid and microdisk cavities are fabricated by standard micro-fabrication processes \cite{50} from a silica-on-silicon wafer, including photolithography, hydrofluoric acid (HF) wet etching, and xenon difluoride (XeF$_2$) dry etching. For the microtoroid, an additional CO$_2$ laser reflow process is performed to decrease the surface roughness and therefore improve the optical $Q$ factors. The diameters of the silicon pedestals for both the microtoroid and microdisk are kept to be relatively small, to increase their mechanical compliance and decrease the mechanical energy dissipation from the cavities to the substrate, and therefore enable higher mechanical \emph{Q} factors. Figures \ref{fig2}(a) and \ref{fig2}(d) show the top-view optical microscope images of the microtoroid and microdisk. Figures \ref{fig2}(b) and \ref{fig2}(e) are the finite element method (FEM) simulated optical field distributions of the fundamental whispering gallery mode (WGM) of the microtoroid and microdisk, where the optical field is confined around the periphery of the microcavities. The measured transmission spectra for one WGM of the microtoroid and microdisk $\sim$1550~nm are shown in Figs. \ref{fig2}(c) and \ref{fig2}(f), respectively. We can see mode splittings for both cavities, which are caused by the backscattering from the surface roughness of the cavities. From the double-peak Lorentzian fittings of the transmission spectra (the red curves), we can derive the intrinsic optical \emph{Q} factor to be about 10$^7$ for the microtoroid, and 4$\times10^5$ for the microdisk. The optical \emph{Q} factors allow 3~dB bandwidths of 16.8~MHz and 419~MHz for the microtoroid and microdisk, respectively, considering the frequency dependence of the shot noise (Eq. (1)).


The measurement setup for ultrasound sensing using the microcavities is shown in Fig. \ref{fig2}(g), with a scanning electron microscope (SEM) image of the microtoroid included. Light from a tunable narrow-linewidth fiber laser in the 1550~nm wavelength band is coupled into the WGMs of the microcavity, through a tapered fiber \cite{51}. The transmitted light from the tapered fiber is detected by a photodetector, and monitored by an oscilloscope to obtain the transmission spectra. The principle of ultrasound detection is shown in the inset of Fig. \ref{fig2}(g). When an ultrasonic wave is applied to the sensor, it can drive the mechanical motion of the cavity and induce a change in the cavity circumference or the taper-cavity coupling strength. Both translate into an amplitude modulation of the intracavity field, which can be optically readout. In our experiment, we use a proportional-integral-derivative (PID) controller to lock the laser wavelength on the side of the optical mode with a detuning where the transmission has the largest slope, to optimize the dispersive transduction of ultrasound signal. The mechanical spectrum of the microcavity is measured with an electronic spectrum analyzer (ESA). The ultrasound signal is produced by an ultrasonic transducer. For experimental convenience, the angle between the incident ultrasound and the disk surface is kept to be $\sim$30$^{\circ}$. We use a function generator to apply a single-frequency sinusoidal voltage to the transducer to measure the single-frequency response of the sensor, and use a vector network analyzer (VNA) to sweep the frequency of the applied ultrasonic wave to obtain the system response of the ultrasound sensor. 

To obtain the sensitivity in a broad frequency range, two piezoelectric ultrasonic transducers with center frequencies at 1~MHz and 5~MHz are used. Considering the attenuation of ultrasonic waves in air, the relation between the ultrasound pressure at the sensor (\emph{P}$_{\rm{sensor}}$) and that at the ultrasonic transducer (\emph{P}$_{\rm{PZT}}$) is \emph{P}$_{\rm{sensor}}$($\omega$) = \emph{e}$^{-\alpha(\omega)\emph{d}}$\emph{P}$_{\rm{PZT}}$($\omega$), where \emph{d} is the distance between the ultrasonic transducer and the sensor, which is kept to be $\sim$1~cm in our experiment. $\alpha(\omega)$ is the frequency dependent acoustic attenuation coefficient, which is obtained from the Stokes-Kirchhoff formula \cite{stokes,52}

\begin{equation}
\alpha(\omega) = \frac{\omega^2}{2\rho v^3}[ \frac{4}{3}\eta’ +(\frac{1}{C_v}-\frac{1}{C_p})]
\end{equation}
where $\rho$ is the density, \emph{v} is the speed of sound, $\eta’$ is the dynamic viscosity coefficient, \emph{C}$_v$ and \emph{C}$_p$ are the specific heat capacities at constant volume and constant pressure, respectively. From this formula, we can see that the absorption loss is proportional to the square of frequency. For example, $\alpha$ is 0.30~dB/cm and 122~dB/cm for ultrasonic waves of 0.5~MHz and 10~MHz, respectively. This frequency dependent absorption loss makes high frequency ultrasound sensing in air challenging. The generated ultrasound pressures at different frequencies are calibrated using a hydrophone. We measure the ultrasound pressure generated by the transducer in water at different frequencies with the hydrophone, and then derive the pressure in air, taking into account the acoustic impedance mismatch, \emph{P}$_{\rm{air}}$ = $\emph{P}_{\rm{water}}\cdot\frac{\emph{Z}_{\rm{air}}}{\emph{Z}_{\rm{water}}}$ = $\frac{\emph{P}_{\rm{water}}}{3580}$, where $Z$ = $\rho$\emph{v} is the acoustic impedance of the material.

\section{Results}

\subsection{Ultrasound detection with the microtoroid}

\begin{figure}[t!]
\begin{center}
\includegraphics[width=8.6cm]{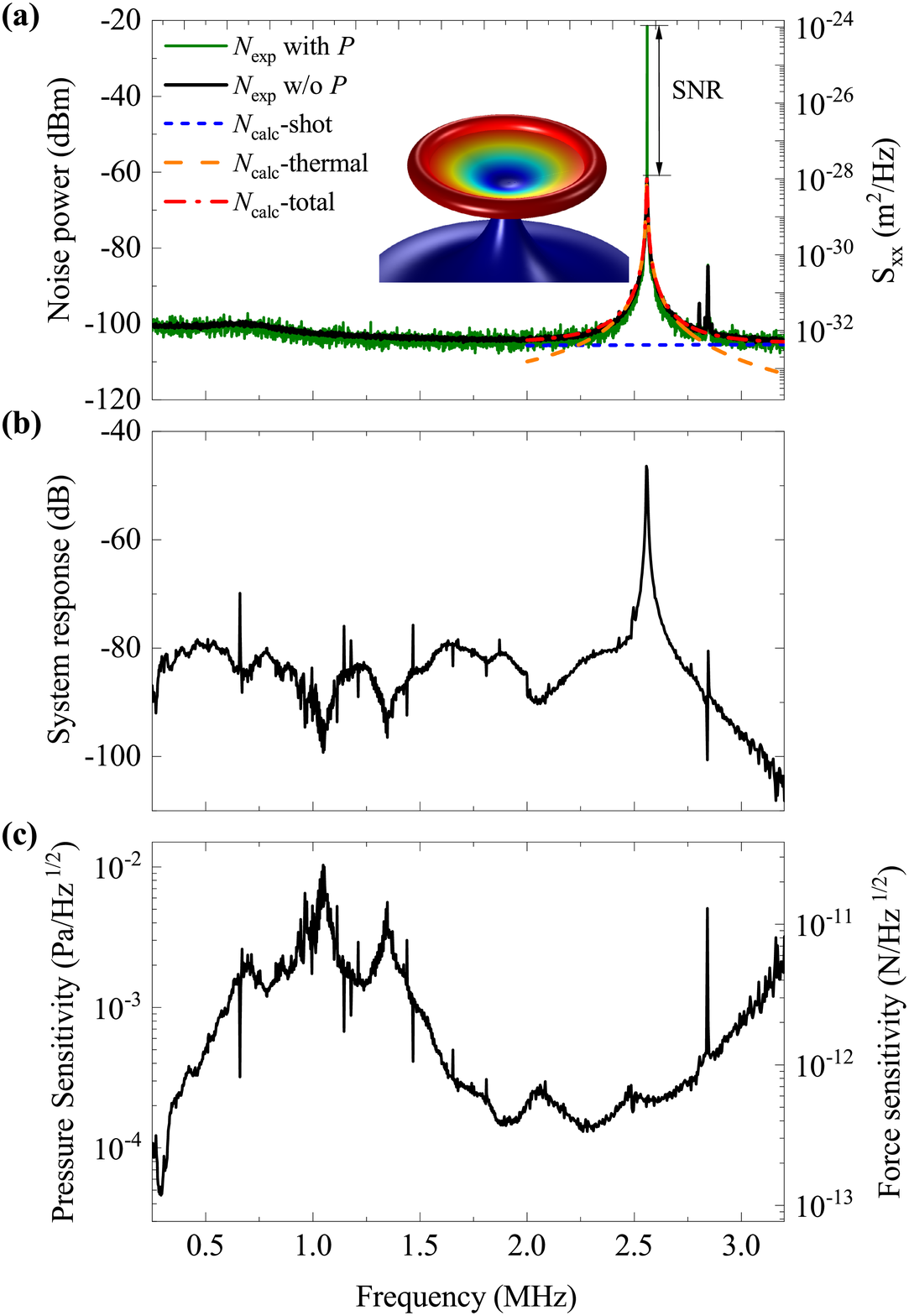}
\end{center}
\caption{(a) Noise power spectrum (black solid curve) and the response of the microtoroid (green solid curve) to an ultrasound at 2.56~MHz, with an SNR of 41.39~dB. The orange dashed, blue short-dashed, and red dash-dotted curves are the calculated thermal noise, shot noise, and total noise, respectively, in the frequency range of 2-3.2~MHz. The inset shows the simulated mode profile of the first-order flapping mode. (b) System response of the microtoroid versus the ultrasound frequency. (c) Derived pressure (left axis) and force (right axis) sensitivity spectra of the microtoroid, with a peak pressure (force) sensitivity of 46~\textmu Pa Hz$^{-1/2}$ (118~fN Hz$^{-1/2}$) at 0.29~MHz.}
\label{fig3}
\end{figure}

We first measure the ultrasound sensitivity of the microtoroid cavity. In order to enable a thermal-noise-limited sensing, but avoid the radiation pressure induced self-sustained mechanical oscillation \cite{radiation1,radiation2}, we keep the input power to be $\sim$10~\textmu W. At this input power, the laser noise is dominated by shot noise in the MHz frequency range. The noise power spectrum measured with the ESA is shown in the black solid curve in Fig. \ref{fig3}(a), in which we can see a mechanical resonance at $\Omega/2\pi$ = 2.56~MHz. This corresponds to the first-order flapping mode, with its mode profile shown in the inset. The thermal noise, shot noise, and total noise in the frequency range of 2-3.2~MHz, calculated from Eq. (1), are shown in the orange dashed, blue short-dashed, and red dash-dotted curves in Fig. \ref{fig3}(a). The corresponding displacement power spectral density (\emph{S}$_{xx}$) of the sensor is shown on the right axis of Fig. \ref{fig3}(a). From the linewidth of the resonance, we can obtain the mechanical \emph{Q} factor of this mode to be $\sim$700. Regarding the mechanical damping of the mode, we theoretically calculate the contribution from the gas damping, which consists of the drag force damping and squeeze-film damping \cite{gas damping}. The drag force damping rate is calculated to be $\gamma_{\rm{drag}}=\mu l_{\rm{drag}}/m=561$~Hz, with $\mu=1.8\times 10^{-5}$~kg m$^{-1}$ s$^{-1}$ being the coefficient of viscosity, $l_{\rm{drag}}=16r\sqrt{m/M}=0.376$~mm is the geometry dependent characteristic length, with $r=28.5$~\textmu m being the radius of the toroid. $m=12.1$~ng is the effective mass, and $M=17.8$~ng is the real mass of the toroid. For the squeeze-film damping, the characteristic length $l_{\rm{squeeze}}=\frac{3\pi r^4}{2h^3}\sqrt{m/M}=0.32$~mm, with $h=20$~\textmu m being the distance between the cavity disk and the substrate. Thus, the squeeze-film damping rate is $\gamma_{\rm{squeeze}}=\mu l_{\rm{squeeze}}/m$=482~Hz. The measured mechanical damping rate of $\gamma_{\rm{meas}}=2\pi\times3.66$~kHz is much larger than both $\gamma_{\rm{drag}}$ and $\gamma_{\rm{squeeze}}$, and therefore we believe the main damping rate of the microtoroid is induced by the intrinsic damping of the cavity structure. 

When we apply an ultrasound signal with a pressure \emph{P}$_{\rm{applied}}$ = 132.2~mPa to the microtoroid sensor at 2.56~MHz, we obtain an SNR of 41.39~dB, measured with a resolution bandwidth $\Delta$$f$ = 20~Hz, as shown in the green solid curve in Fig. \ref{fig3}(a). The sensitivity at 2.56~MHz can be calculated by the following equation
\begin{equation}
P_{min}(\Omega) = P_{applied}(\Omega)\sqrt{\frac{1}{\rm{SNR}}\cdot\frac{1}{\Delta f}} \sim 252\; \upmu \rm{ Pa Hz^{-1/2}}
\end{equation}

Using the parameters in our experiment, the relative pressure difference ratio of the sensor is obtained to be \emph{r} = 1.17 at 2.56~MHz, which is larger than 1, due to the substrate reflection enhanced ultrasound pressure at the microtoroid. Considering the incident angle $\theta$ = 30$^{\circ}$ of the ultrasound, the spatial overlap $\zeta$ between the incident ultrasound and the first-order flapping mode of the microtoroid is 0.39. We can then derive the theoretical sensitivity at this frequency to be 41~\textmu Pa Hz$^{-1/2}$. The difference between the experimental and theoretical sensitivities could be resulting from the misalignment of the ultrasonic transducer to the microtoroid sensor.

We then use a network analyzer to drive the ultrasonic transducer to obtain the system response of our sensor, for ultrasonic waves with different frequencies. In order to obtain the sensor response in a broad frequency band, we use two ultrasonic transducers, with center frequencies at 1~MHz and 5~MHz, respectively. System response in the frequency range of 0.25-3.2~MHz is obtained, as shown in Fig. \ref{fig3}(b). The lower frequency limit of 0.25~MHz is not intrinsic, but rather limited by the low pressure of the ultrasound produced by the transducer. The upper limit of 3.2~MHz is introduced by the larger attenuation of air at higher frequencies ($\alpha$ = 12.5~dB/cm at 3.2~MHz). It can be seen that the response of the sensor around the resonance frequency of 2.56~MHz is significantly enhanced, due to the high mechanical \emph{Q} factor of the mode and the large spatial overlap between the mode displacement and the ultrasonic wave. Other peaks in the frequency band correspond to other mechanical modes of the toroid or the tapered fiber. These modes do not reach thermal-noise-dominant regime, and are therefore not seen in the noise power spectrum in Fig. \ref{fig3}(a). 

From the system response \emph{S}($\omega$) and the noise power spectral density \emph{N}($\omega$), combined with the sensitivity \emph{P}$_{\rm{min}}$($\Omega$) at $\Omega$/2$\pi$ = 2.56 MHz, we can derive the sensitivity over the entire frequency range:
\begin{equation}
P_{\rm{min}}(\omega) = P_{\rm{min}}(\Omega)\frac{P_{\rm{applied}}(\omega)}{P_{\rm{applied}}(\Omega)}\sqrt{\frac{N(\omega)}{N(\Omega)}\cdot\frac{S(\Omega)}{S(\omega)}}
\end{equation}
where \emph{P}$_{\rm{applied}}$($\omega$) is the applied ultrasound pressure to the sensor at different frequencies. The pressure sensitivity in the frequency range of 0.25-3.2~MHz is shown on the left axis of Fig. \ref{fig3}(c). Multiplying the sensor area, we can obtain the force sensitivity of the sensor, as shown on the right axis of Fig. \ref{fig3}(c). A peak pressure (force) sensitivity of 46~\textmu Pa Hz$^{-1/2}$(118~fN Hz$^{-1/2}$) is achieved at~0.29 MHz. Around the mechanical resonance frequency, thermal-noise-limited pressure (force) sensitivity is reached to 130-475~\textmu Pa Hz$^{-1/2}$(0.34-1.21~pN Hz$^{-1/2}$) in the frequency range of 2.24-2.84~MHz. In the whole frequency range of 0.25-3.2~MHz, the pressure (force) sensitivity is better than 10~mPa Hz$^{-1/2}$(26.4~pN Hz$^{-1/2}$). In terms of force sensitivity, the peak sensitivity of 118~fN Hz$^{-1/2}$ of our sensor is about three times better than that in Ref. \cite{45}.

\begin{figure}[t!]
\begin{center}
\includegraphics[width=8.6cm]{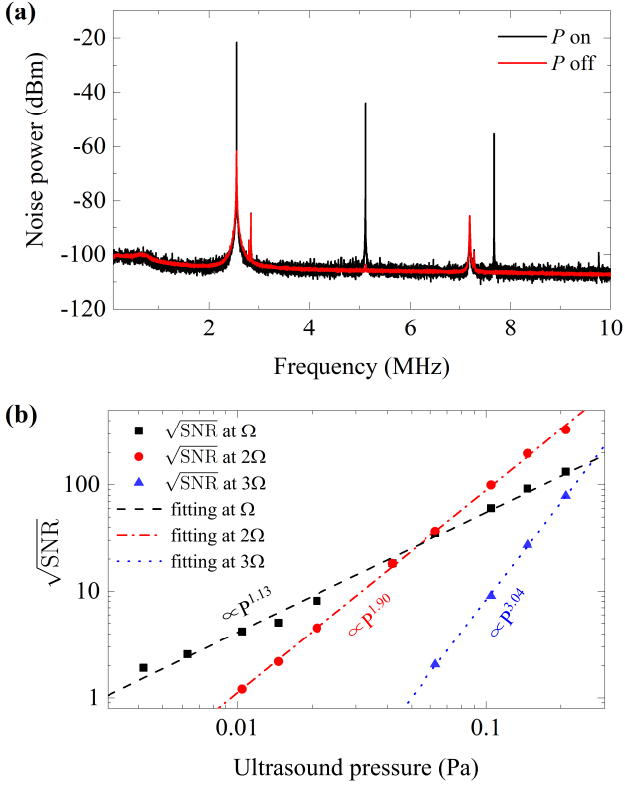}
\end{center}
\caption{(a) Noise power spectrum with (black curve) and without (red curve) applying a single-frequency ultrasound at 2.56~MHz. (b) $\sqrt{\rm{SNR}}$ versus the ultrasound pressure, at $\Omega$ (black squares), 2$\Omega$ (red circles), and 3$\Omega$ (blue triangles) frequencies, with the black dashed line, red dash-dotted line, and blue dotted line representing the corresponding fitting results.}
\label{fig4}
\end{figure}

When we apply an ultrasound at the mechanical resonance frequency $\Omega$/2$\pi$ = 2.56 MHz, in addition to a response peak at 2.56~MHz, we also observe responses at the second- and third-order mechanical sidebands, as shown in Fig. \ref{fig4}(a). It can be seen that, when the ultrasound pressure is applied, three peaks at $\Omega$, 2$\Omega$, and 3$\Omega$ frequencies appear in the noise power spectrum. At the input power of 10~\textmu W, the radiation pressure force induced mechanical oscillations can be neglected, and the higher order mechanical sidebands are induced by the nonlinear transduction. Since the cavity mode is a Lorentzian lineshape, the optical readout signal for displacement is a harmonic oscillation only for a small displacement. In the large displacement case, the readout signal becomes anharmonic. Previous works have experimentally studied the intensities of the high order mechanical sidebands as a function of the optical power \cite{53,54,55,56,57}, and theoretically studied the dependence of the high order mechanical sidebands on the displacement \cite{58}. Here we experimentally study the intensities of the high order mechanical sidebands with different mechanical displacements, driven by an ultrasonic wave. Figure \ref{fig4}(b) shows the measured SNRs at the first- (black squares), second- (red circles), and third- (blue triangles) order sidebands, respectively. By performing exponential fittings to these experimental results, we obtain that $\rm{\sqrt{SNR}}$($\Omega$)$\propto$\emph{P}$^{\rm{1.13}}$, $\rm{\sqrt{SNR}}$(2$\Omega$)$\propto$\emph{P}$^{\rm{1.90}}$, $\rm{\sqrt{SNR}}$(3$\Omega$)$\propto$\emph{P}$^{\rm{3.04}}$. In the following we theoretically study the dependence of the SNR on the mechanical displacement at the three sidebands.

From the equation of motion of the cavity mode $\dot a$ = $-\kappa a/2$ + \emph{i}$\Delta$\emph{a} + $\sqrt{\kappa_{\rm{e}}\emph{s}}$, we can obtain the intracavity photon number to be \emph{n}$_{\rm{c}}$ = $\frac{4\kappa_{\rm{e}}\emph{s}}{\kappa^2}$$\frac{1}{1+\delta^2}$, with \emph{s} being the number of photons injected into the microcavity per unit time, and $\kappa$ = $\kappa_0$ + $\kappa_{\rm{e}}$ being the total energy decay rate of the cavity mode. $\kappa_0$ is the intrinsic decay rate of the cavity mode, and $\kappa_{\rm{e}}$ is the coupling rate with the tapered fiber. $\delta$ = $2\Delta$/$\kappa$ is the dimensionless detuning, with $\Delta$ = $\omega-\omega_0$ denoting the frequency detuning between the input light and the cavity mode. Taylor expanding the detuning $\delta$, we can obtain the intracavity photon number to be:

\begin{equation}
n_{\rm{c}} \approx n_{\rm{c}}^{\rm{max}}[c_0(\delta)+c_1(\delta)u+c_2(\delta)u^2+c_3(\delta)u^3]
\end{equation}
where $n_{\rm{c}}^{\rm{max}}$ is the introcavity photon number when $\Delta$ = 0, \emph{c}$_0$($\delta$) = 1/(1+$\delta$$^2$), \emph{c}$_{\rm{n}}$($\delta$) = $\frac{1}{\emph{i}!}$$\frac{\emph{d}^{\rm{i}}}{\emph{d}\delta^{\rm{i}}}$\emph{c}$_0$($\delta$). \emph{u} = 2\emph{G}\emph{x}/$\kappa$ represents the normalized frequency shift of the cavity mode caused by the mechanical displacement. Using the input-output relation $a_{\rm{out}}=\sqrt{s}-\sqrt{\kappa_{\rm{e}}}a$, we can obtain the photocurrent arriving at the photodetector at a certain detuning:

\begin{equation}
\begin{aligned}
Z = |a_{\rm{out}}|^2 \approx&s-\frac{4{\kappa_{\rm{e}}}{\kappa_0}s}{\kappa^2}
[c_0(\delta)+c_1(\delta) \frac{2G}{\kappa}x+c_2(\delta) (\frac{2G}{\kappa})^2x^2\\
&+c_3(\delta) (\frac{2G}{\kappa})^3x^3]
\label{f2}
\end{aligned}
\end{equation}

Expressing the displacement of the cavity caused by the ultrasonic wave with \emph{x} = \emph{x}$_0$cos($\Omega$\emph{t}), we can obtain the following coefficients of the photocurrent for DC, $\Omega$, 2$\Omega$, and 3$\Omega$ frequency components:
\begin{subequations}
	
	\begin{equation}
		Z_{\rm{DC}} = s - \frac{4{\kappa_{\rm{e}}}{\kappa_0}s}{\kappa^2}c_0(\delta)
	\end{equation}
	
	\begin{equation}
		Z_{\Omega} = - \frac{4{\kappa_{\rm{e}}}{\kappa_0}s}{\kappa^2}c_1(\delta) \frac{2G}{\kappa}x_0\rm{cos}(\Omega t)
	\end{equation}
	
	\begin{equation}
		Z_{2\Omega} = - \frac{4{\kappa_{\rm{e}}}{\kappa_0}s}{\kappa^2}[c_2(\delta) (\frac{2G}{\kappa})^2\frac{x_0^2}{2}\rm{cos}(2\Omega t)]
	\end{equation}
	
	\begin{equation}
		Z_{3\Omega} = - \frac{4{\kappa_{\rm{e}}}{\kappa_0}s}{\kappa^2}[c_3(\delta) (\frac{2G}{\kappa})^3\frac{x_0^3}{4}\rm{cos}(3\Omega t)]
	\end{equation}
	
\end{subequations}

As the amplitude of the mechanical displacement \emph{x}$_0$ is proportional to the ultrasound pressure \emph{P}, we can obtain the dependence of SNR on the ultrasound pressure \emph{P} at the three mechanical sidebands: $\rm{\sqrt{SNR}}$($\Omega$)$\propto$\emph{P}, $\rm{\sqrt{SNR}}$(2$\Omega$)$\propto$\emph{P}$^{\rm{2}}$, $\rm{\sqrt{SNR}}$(3$\Omega$)$\propto$\emph{P}$^{\rm{3}}$, respectively, which explains our experimental results well. Measuring the combination of SNRs at all the mechanical sidebands has the potential to extend the dynamic range of displacement sensing \cite{53}.

\subsection{Ultrasound detection with the microdisk}

\begin{figure*}[t!]
\begin{center}
\includegraphics[width=17.2cm]{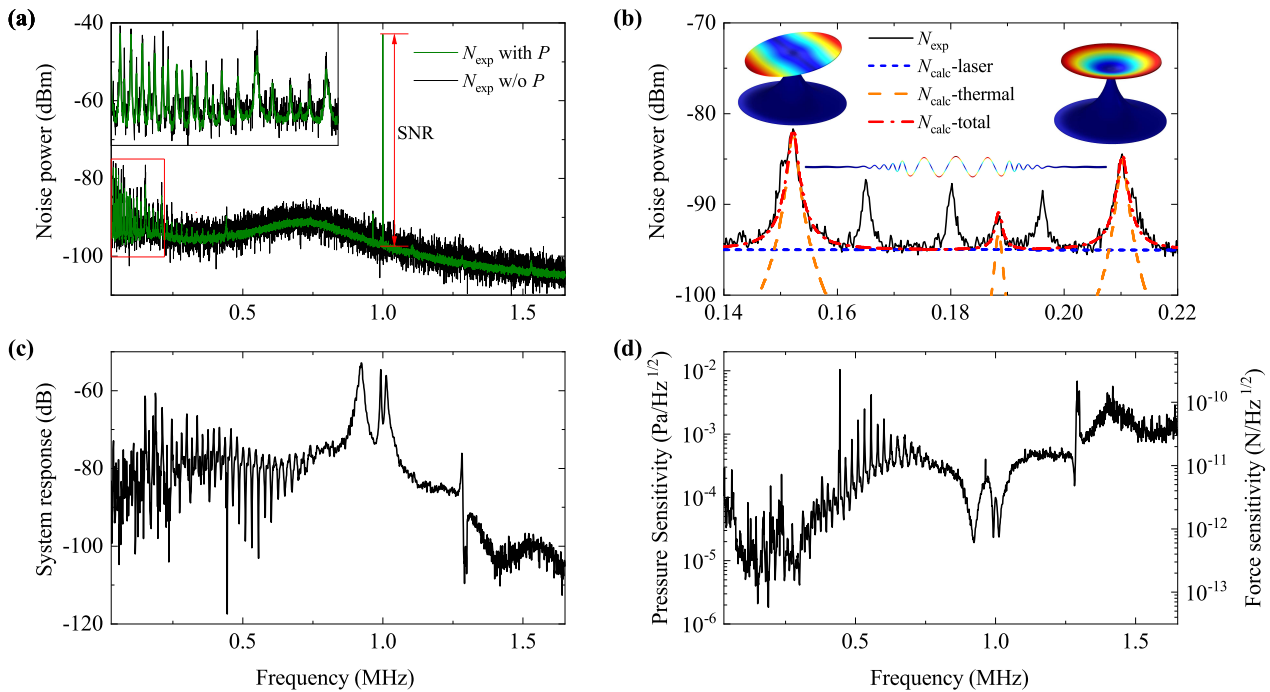}
\end{center}
\caption{(a) Noise power spectra with (green solid curve) and without (black solid curve) applying an ultrasound at 1~MHz, showing an SNR of 54.43~dB. The inset shows a zoomed-in noise power spectrum in the red rectangle in (a), showing multiple mechanical resonances of the tapered fiber and the microdisk in the tens-to-hundreds of kHz range. (b) Zoomed-in noise power spectrum of (a) in the frequency range of 140-220~kHz, with the black solid curve representing the measured noise power spectrum. The orange dashed, blue short-dashed, and red dash-dotted curves are the calculated thermal noise, laser noise, and total noise, respectively. The insets show the simulated mode profiles of the tilting mode of the disk at 152~kHz, first-order flapping mode at 210~kHz of the disk, and one string mode of the tapered fiber at 189~kHz. (c) System response of the microdisk sensor versus the ultrasound frequency. (d) Derived pressure (left axis) and force (right axis) sensitivity spectra of the microdisk, with a peak pressure (force) sensitivity of 1.83~\textmu Pa Hz$^{-1/2}$ (58~fN Hz$^{-1/2}$) at 189~kHz.}
\label{fig5}
\end{figure*}

In order to improve the ultrasound sensitivity in the kHz frequency range, we then use a microdisk with a diameter of 200~\textmu m to perform the ultrasound sensing. To enable a thermal-noise-dominant sensing, the input optical power is kept to be $\sim$100~\textmu W (limited by the saturation power of the photodetector), which is well below the threshold of radiation pressure induced self-sustained mechanical oscillation, due to the relatively low optical $Q$ factor of the microdisk. The measured noise power spectrum of the microdisk in 30~kHz-1.65~MHz is shown in the black curve in Fig. \ref{fig5}(a). There exhibits a wide noise peak around 700~kHz, which is possibly the characteristic relaxation oscillation noise of the fiber laser \cite{relaxation oscillation}. This intensity noise of the laser can be suppressed by using a homodyne detection method. When applying an ultrasound signal with a pressure \emph{P}$_{\rm{applied}}$ = 287.6~mPa to the microdisk sensor at 1~MHz, we obtain an SNR of 54.43~dB at this frequency, as shown in the green solid curve in Fig. \ref{fig5}(a). Using the SNR and resolution bandwidth of 30~Hz, the sensitivity at 1~MHz is derived from Eq. (3) to be 99.7~\textmu Pa Hz$^{-1/2}$.

In addition to the mechanical resonances of the microdisk, multiple resonance peaks are also observed from tens-to-hundreds of kHz range, as shown in the inset of Fig. \ref{fig5}(a), which is a zoomed-in noise power spectrum in 30-220~kHz. We conclude from FEM simulations that these peaks correspond to the mechanical modes (string modes of different orders) of the tapered fiber which can be optically readout through dissipative coupling between the taper and the microdisk. Figure \ref{fig5}(b) is a further zoom-in of the noise power spectrum of Fig. \ref{fig5}(a), in the range from 140-220~kHz. In this range, multiple mechanical resonances appear, including the tilting mode and first-order flapping mode of the microdisk, and six of the taper modes (with two of them overlapped with the two disk modes). The mode profiles are obtained from FEM simulations and shown in the insets of Fig. \ref{fig5}(b), for the tilting mode (at $\sim$152~kHz) and first-order flapping mode (at $\sim$210~kHz) of the microdisk, and one string mode of the tapered fiber (at $\sim$189~kHz). For these three modes, we can calculate the thermal noise, laser noise, and total noise from Eq. (1), plotted in the orange dashed, blue short-dashed, and red dash-dotted curves in Fig. \ref{fig5}(b), respectively. From the linewidths of the resonances, we can obtain the mechanical \emph{Q} factors to be 100, 130, and 108 for the three modes.





We then measure the system response of the microdisk ultrasound sensor in 30~kHz-1.65~MHz range with a network analyzer, with the result shown in Fig. \ref{fig5}(c). We can see that the ultrasound response is enhanced by the mechanical resonances of the microdisk and tapered fiber, and the lower frequency limit is extended to 30~kHz, which is much smaller than that using the microtoroid (0.25~MHz). The upper frequency limit of 1.65~MHz is also limited by the larger attenuation of air at higher frequencies. The response peak at 924~kHz (1.282~MHz) corresponds to the third-order crown mode (second-order flapping mode) of the microdisk. The two peaks at 992~kHz and 1.012~MHz are split modes of the fourth-order crown mode of the microdisk, possibly caused by the fabrication induced slight asymmetry of the microdisk. The ultrasound response is significantly enhanced at these four modes, but they are not observed in the noise power spectrum in Fig. \ref{fig5}(a), as the noise of the sensor is dominated by the laser noise at these frequencies. 

With the noise power spectrum, system response, and the ultrasound sensitivity at 1~MHz, we can derive the sensitivity in the frequency range of 30~kHz-1.65~MHz from Eq. (4), as shown in Fig. \ref{fig5}(d), with the corresponding force sensitivity shown on the right axis. A peak pressure (force) sensitivity of 1.83~\textmu Pa Hz$^{-1/2}$ (58~fN Hz$^{-1/2}$) is achieved at 189~kHz, corresponding to a mechanical mode of the tapered fiber. This pressure sensitivity is higher than any cavity optomechanical ultrasound sensor reported so far. The pressure (force) sensitivity is 2.09~\textmu Pa Hz$^{-1/2}$ (66~fN Hz$^{-1/2}$) for the tilting mode (at 152~kHz), and 4.95~\textmu Pa Hz$^{1/2}$ (165~fN Hz$^{-1/2}$) for the first-order flapping mode (at 210~kHz) of the disk. It is worth noting that the sensitivities of the tilting and first-order flapping modes are higher than their theoretical values. For instance, for the flapping mode, using the simulated pressure difference ratio $r=0.058$, and spatial overlap $\zeta=0.3$, its theoretical sensitivity is derived to be $\sim126$~\textmu Pa Hz$^{-1/2}$, which is about 25 times larger than the experimental sensitivity. This is because there are taper modes superposed to these two disk modes (as shown in Fig. \ref{fig5}(b)), and therefore the sensitivities at these frequencies are enhanced by the mechanical resonances of the tapered fiber. Due to the larger sensing area of the tapered fiber, the taper modes provide higher sensitivities than the disk modes. Note that the sensitivity at 0.29~MHz for the microtoroid ultrasound sensor could also be enhanced by the mechanical resonances of the tapered fiber, which explains why the highest sensitivity occurs at 0.29~MHz, even though there is no mechanical mode of the toroid.



\section{Conclusions}

We have demonstrated air-coupled high-sensitivity MHz frequency ultrasound detection based on high $Q$ WGM microcavities. Using an on-chip microtoroid cavity with a principal diameter of 57~\textmu m, high optical $Q$ factor of $10^7$, and mechanical \emph{Q} factor of 700, we have extended the air-coupled ultrasound sensing into the MHz frequency range, and achieved sensitivities of 46~\textmu Pa Hz$^{-1/2}$-10~mPa Hz$^{-1/2}$ in 0.25-3.2~MHz frequency range, with a thermal noise limited range of 0.6~MHz. 
In addition, we have observed the second- and third-order mechanical sidebands when driving the sensor with an ultrasound at the mechanical resonance frequency, and the measured intensities at three mechanical sidebands are consistent with our theoretical results. This nonlinear transduction provides a way to extend the dynamic range of displacement sensing. Furthermore, in order to improve the sensitivity in the kHz frequency range, we have used an on-chip microdisk cavity with a diameter of 200~\textmu m and achieved sensitivities of 1.83~\textmu Pa Hz$^{-1/2}$-10.4~mPa Hz$^{-1/2}$ in 30~kHz-1.65~MHz range. The peak sensitivity is the record sensitivity of cavity optomechanical ultrasound sensors reported so far.

The ultrasound sensitivity can be further improved by using a larger and thinner cavity, realizing a larger pressure difference ratio \emph{r} by designing the structure, and optimizing the incident angle of the ultrasound. The use of mechanical modes with stronger optomechanical coupling coefficient \cite{17} and squeezed light \cite{31} can reduce shot noise and expand the thermal-noise-dominant regime. Integrated waveguide-coupled microcavities \cite{59} and on-chip arrays of sensors can be designed in the future, for photoacoustic imaging and spectroscopy \cite{2,3,4}. This work broadens the frequency range of ultrasound detection in air, which is of great significance for applications in gas photoacoustic spectroscopy, and non-contact ultrasonic medical imaging, etc. The photoacoustic signal near the resonance frequency has an enhanced response, which can be applied to high-sensitivity biomedical measurements \cite{48,49}.

\begin{acknowledgments}
We thank the funding support from The National Key Research and Development Program of China (2021YFA1400700), the National Natural Science Foundation of China (NSFC) (91950118, 12174438, 11934019), and the basic frontier science research program of Chinese Academy of Sciences (ZDBS-LY- JSC003).
\end{acknowledgments}

\end{document}